# Nonlinear force-free models for the solar corona

## I. Two active regions with very different structure

S. Régnier and E. R. Priest

School of Mathematics and Statistics, University of St Andrews, Fife, KY16 9SS, UK  
e-mail: `(stephane, eric)@mcs.st-andrews.ac.uk`

**ABSTRACT**

*Context.* With the development of new instrumentation providing measurements of solar photospheric vector magnetic fields, we need to develop our understanding of the effects of current density on coronal magnetic field configurations.
*Aims.* The object is to understand the diverse and complex nature of coronal magnetic fields in active regions using a nonlinear force-free model.
*Methods.* From the observed photospheric magnetic field we derive the photospheric current density for two active regions: one is a decaying active region with strong currents (AR8151), and the other is a newly emerged active region with weak currents (AR8210). We compare the three-dimensional structure of the magnetic fields for both active region when they are assumed to be either potential or nonlinear force-free. The latter is computed using a Grad-Rubin vector-potential-like numerical scheme. A quantitative comparison is performed in terms of the geometry, the connectivity of field lines, the magnetic energy and the magnetic helicity content.
*Results.* For the old decaying active region the connectivity and geometry of the nonlinear force-free model include strong twist and strong shear and are very different from the potential model. The twisted flux bundles store magnetic energy and magnetic helicity high in the corona (about 50 Mm). The newly emerged active region has a complex topology and the departure from a potential field is small, but the excess magnetic energy is stored in the low corona and is enough to trigger powerful flares.

## 1. Introduction

The coronal magnetic field is complex in nature. Coronal observations have shown the diversity and complexity of active region magnetic fields evidenced by filaments (Priest et al. 1989; Martin 1998), sigmoids (Rust & Kumar 1996; Canfield et al. 1999), and flare sites (e.g., Masuda et al. 1995). Many models have been developed in order to determine the links between the complexity of the magnetic field and flaring activity (see review by Priest & Forbes 2002). In a coronal environment dominated by the magnetic field (low plasma $\beta$), the main source responsible for the complexity of the field is the existence of electric currents along field lines. The currents originate either from below the photospheric surface (flux emergence) or from the horizontal velocity fields on the photosphere (convective motions).

In addition to the study of active regions, the complexity of the coronal magnetic field has been studied in the Quiet Sun (Close et al. 2004) and for the global magnetic field (Mackay & van Ballegooijen 2006; Riley et al. 2006; Maclean et al. 2006a,b).

*Send offprint requests to*: S. Régnier

The electric current density in solar magnetic configurations was first measured by Rayrole & Semel (1970) and Krall et al. (1982) from spectro-polarimetric observations in strong field regions. The authors showed how the vertical current density can be derived from the measurement of the three components of the magnetic field on the photosphere. The distribution of current density was found to be nonuniform and with a large spread of values (even in sign) in one polarity showing that active region magnetic fields can store magnetic energy and have a complex geometry (e.g., twisted flux bundles, sheared arcades). Observations with high spatial resolution (such as SOHO/EIT, TRACE, Yohkoh/SXT) have shown that non-potential models of the magnetic fields fit the observations better than potential models which confirms the existence of currents in active regions. The missing link is currently our lack of understanding of the effects of nonuniformly distributed photospheric currents on three-dimensional coronal configurations.

Since the first attempts to reconstruct the coronal magnetic field from observations (Schmidt 1964; Semel 1967), there has been a growing interest in determining the coronal magnetic field from photospheric measurements, especially with the de-velopment of spectropolarimeters and of numerical techniques. The basic idea is to suppose the magnetic configuration of an active region is in an equilibrium state between magnetic, pressure and gravity forces. A special case of this magnetohydrostatic equilibrium is obtained for coronal conditions where the pressure and the gravity can be neglected. We then have a so-called force-free field equilibrium satisfying:

$$\mathbf{j} \wedge \mathbf{B} = \mathbf{0}, \tag{1}$$

where $\mathbf{B}$ is the magnetic field and $\mathbf{j}$ is the current density (the solenoidal equation should also be satisfied). Three different types of solution of Eqn.(1) are commonly considered:

(i) a potential field for which the current density is zero everywhere in the coronal volume. The derived magnetic field is relatively easy to compute nowadays and several methods with different numerical schemes and different boundary conditions have been developed. The potential field corresponds to the minimum energy state that a magnetic configuration can reach with the same normal magnetic component on the boundaries (see e.g. Schmidt 1964; Semel 1967; Altschuler & Newkirk 1969);

(ii) a linear force-free field assumes that the current density is proportional to the magnetic field with a constant $\alpha$ ($\alpha$ being the same at each location in the coronal volume). Several numerical methods have been developed and interesting results have been derived. The magnetic field obtained following this assumption is a minimum energy state for a given total relative magnetic helicity (see e.g. Nakagawa & Raadu 1972; Chiu & Hilton 1977; Alissandrakis 1981; Gary 1989, for the most popular techniques);

(iii) a nonlinear force-free (*nlff*) field assumes that the current density is proportional to the magnetic field with a constant of proportionality ($\alpha$) that varies with space. The *nlff* field also satisfies the additional constraint that $\alpha$ is constant along each field line. More challenging in terms of computation, several *nlff* methods have been developed and have been applied successfully to solar active regions (see review in Amari et al. 1997; Jiao et al. 1997; Wiegelmann 2004; Schrijver et al. 2006; Régnier 2007).

It has been known since the early 80s that the vertical current density derived from the observed photospheric magnetic field can be positive and negative in one polarity leading to the existence of return current in the corona. This is incompatible with both potential and linear force-free models. The nature of the photosphere can be checked by means of integral properties derived by Molodensky (1969) and Aly (1989). By applying these properties, Metcalf et al. (1995) and Moon et al. (2002) have shown that the photosphere is not force-free but becomes force-free at about 400 km above the photosphere (this height is typically represented by half a pixel in our force-free modeling). Wiegelmann et al. (2006) have developed a preprocessing technique in order to minimize the magnetic forces and torques on the photosphere. The *nlff* approximation is, however, a good approximation for the coronal magnetic field, especially compared to potential and linear force-free fields. The use of a mathematically well-posed problem to solve the *nlff* field ensures that the reconstructed field is force-free even if the transverse field slightly differs from the observed transverse field. It is worth noticing that, when using photospheric magnetic measurements, the magnetohydrostatic assumption should give a better description of the field by allowing electric currents perpendicular to the magnetic field lines (see e.g. Wiegelmann & Neukirch (2006) for a first attempt to reconstruct magnetohydrostatic fields).

In this paper we compare potential field models of active regions with nonlinear force-free models. We re-write Eqn. (1) in terms of $\mathbf{B}$ as follows:

$$\nabla \wedge \mathbf{B} = \alpha \mathbf{B}, \tag{2}$$

and by taking the divergence of the above equation we obtain that

$$\mathbf{B} \cdot \nabla \alpha = 0 \tag{3}$$

where $\alpha$ is defined as the force-free function (e.g. $\alpha = (\nabla \wedge \mathbf{B})_z / B_z$). For a potential field or current-free field, $\alpha$ vanishes everywhere in the considered volume. To compute the *nlff* magnetic field, we extrapolate the photospheric magnetic measurements into the corona using the vector-potential Grad-Rubin-like method (Grad & Rubin 1958) developed by Amari et al. (1997; 1999) and used for solar applications by Bleybel et al. (2002) and Régnier et al. (2002; 2004; 2006). The potential and the *nlff* fields are computed with the same boundary conditions for the normal component of $\mathbf{B}$ and its associated vector potential, $\mathbf{A}$. For the potential field case, the lower boundary condition is given by the observed vertical component of $\mathbf{B}$. For the *nlff* field, we also need to provide the distribution of $\alpha$ derived from the vertical and transverse components of $\mathbf{B}$ at the boundary in one and only one polarity. The latter boundary condition guarantees that we have a mathematically well-posed problem (Sakurai 1981) to solve the *nlff* equations. For the side boundaries, we prescribe the normal component of $\mathbf{B}$ and $\alpha$ to vanish leading to closed boundary conditions. These boundary conditions are suitable for active-region magnetic fields where only the bottom boundary condition is known. Those side boundary conditions imply that the field-of-view should be large enough and the magnetic field should decrease fast enough to be valid. We notice that Amari et al. (2006) and Schrijver et al. (2006) have implemented different side boundary conditions suitable for analytical or semi-analytical solutions.

Even if the current density can be estimated on the photosphere, it is not clear how the change in the current density distribution will affect a coronal magnetic configuration. And it is the aim of our study to understand such modifications in terms of the geometry of field lines, the storage of magnetic energy and the amount of magnetic helicity. In Section 2, we will describe the two active regions and the photospheric magnetic field data used to derive the 3D coronal field. In Section 3, we proceed to a visual inspection of the 3D magnetic configurations as well as a statistical study of the geometrical and magnetic properties of characteristic field lines. And we analyse the magnetic energy and the magnetic helicity budgets of the active regions in Section 4.

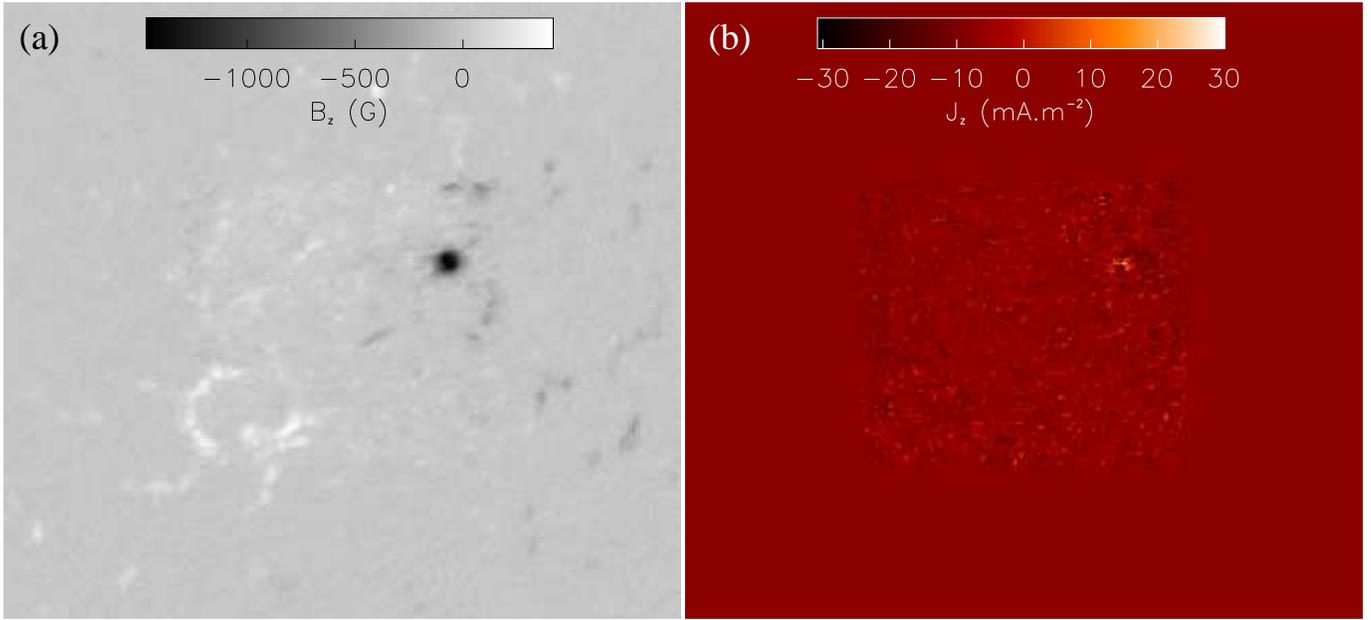

**Fig. 1.** Photospheric distributions for AR 8151: (a) the vertical component of the magnetic field as observed by IVM; (b) the vertical current density as computed from the transverse components. The IVM field-of-view is surrounded by SOHO/MDI magnetic field for $B_z$ and zero values for $J_z$ (see text for details). The observations were recorded on February 11, 1998 at 17:36 UT in a composite field-of-view (SOHO/MDI and IVM) of 330″×300″.

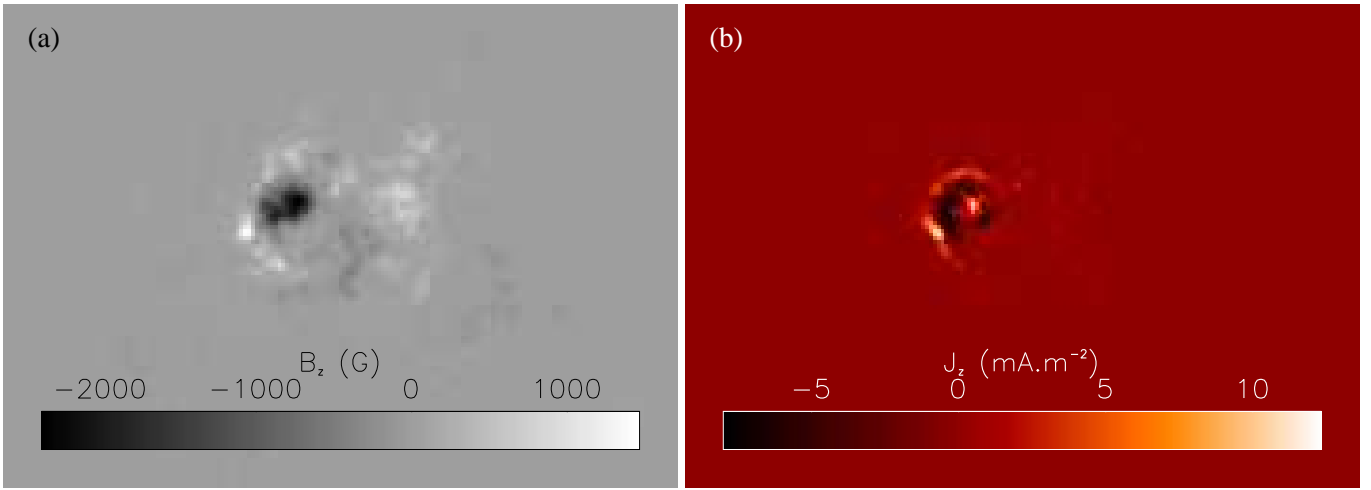

**Fig. 2.** Photospheric distributions for AR 8210: (a) the vertical component of the magnetic field as observed by IVM; (b) the vertical current density as computed from the transverse components. The IVM field-of-view is surrounded by SOHO/MDI magnetic field for $B_z$ and zero values for $J_z$ (see text for details).

## 2. Active regions

### 2.1. Decaying active region

The active region 8151 (AR 8151) was observed on February 10–15, 1998 in the southern hemisphere. The magnetic configuration of this active region has been extensively studied by Régnier et al. (2002) and Régnier & Amari (2004). The authors have found the existence of twisted flux tubes in AR 8151 with different numbers of turns and different handedness.

The photospheric vector magnetic field is provided by MSO/IVM (Mees Solar Observatory/Imaging Vector Magnetograph, Mickey et al. 1996). The observations were performed on February 11, 1998 at 17:36 UT with a field-of-view of 280″ square for a spatial resolution of 1″. The magnetic field distribution (see Fig. 1a) is rather simple: a leading negative sunspot ($B_z \sim -1500$ G) followed by a diffuse positive polarity ($B_z \sim 450$ G). Small scale magnetic features including parasitic polarities are responsible for the complexity of the magnetic field configuration. AR 8151 has reached a stage of its evolution for which the magnetic flux is decaying. In Fig. 1b, we plot the distribution of the vertical current density $J_z$ on the photosphere given by

$$J_{z,phot} = \frac{1}{\mu_0} \left( \frac{\partial B_{y,phot}}{\partial x} - \frac{\partial B_{x,phot}}{\partial y} \right). \quad (4)$$

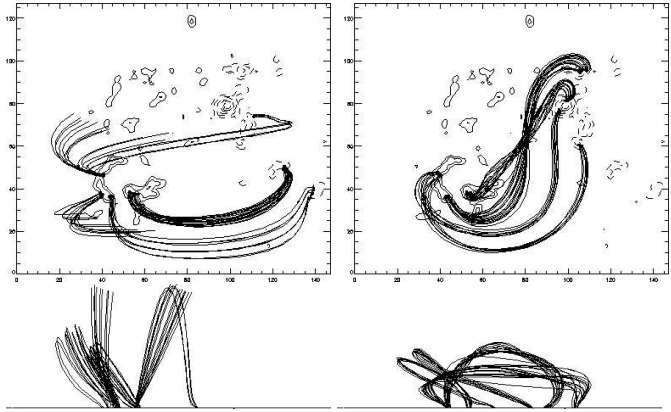

**Fig. 3.** AR 8151 3D magnetic field configurations for the potential field (left) and for the *nlff* field (right). A few particular flux bundles are plotted corresponding to the same footpoints in the positive polarity (solid contours). We note that the current density modifies the geometry of the field lines and their connectivity.

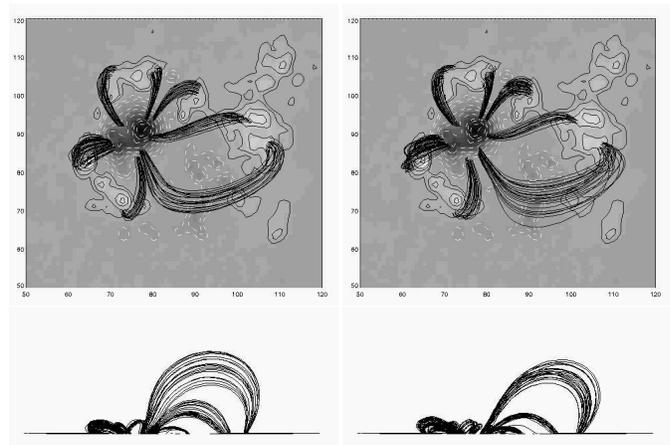

**Fig. 4.** AR 8210 3D magnetic field configurations for the potential field (left) and for the *nlff* field (right). Only a few field lines are plotted characterising the main features of the configurations. There is no evidence of much change of connectivity between both configurations. The background image represents the vertical magnetic field with positive (resp. negative) polarities in white colour or solid contours (resp. black colour or dashed contours).

We estimate the noise level of the magnetic field components following Leka & Skumanich (1999): about 50 G for the vertical component and about 200 G for the transverse components. The $J_z$ distribution ranges from $-30$ to $30$ mA·m$^{-2}$. In the following parts, we will use the values of $\alpha = J_z/B_z$ instead of $J_z$ because the $\alpha$ should be the same at both footpoints of a loop. The $\alpha$ values range from $-1$ to $1$ Mm$^{-1}$.

### 2.2. Newly emerged active region

The active region 8210 (AR 8210) was observed on May 1, 1998 in the southern hemisphere by MSO/IVM. A detailed analysis of the time evolution of AR 8210 has been done in Régnier & Canfield (2006). The authors have especially emphasized the fact that the magnetic configuration exhibits a complex magnetic topology including lots of null points and separatrix surfaces. No twisted flux tubes have been found in the configuration of AR 8210.

The vector magnetic field measured by IVM was recorded on May 1, 1998 at 19:40 UT within a field-of-view of 280″ square and with a resolution of 1″. In order to reduce the noise level on the transverse component, we have averaged the Stokes parameters over 15 minutes (5 consecutive complete observations with a time cadence of 3 minutes). The noise level on the vertical component is reduced to 30 G and on the transverse components to 70 G. As seen in Fig. 2a, the distribution of the vertical component on the photosphere contains a strong negative sunspot surrounded by multiple positive polarities. The strongest positive polarity is located on the south-east side of AR 8210 and the weakest and more diffuse positive polarity is on the west side (also associated with a weak diffuse negative polarity). In Fig. 2b, we plot the distribution of $J_z$ derived from the transverse magnetic field components from Eqn. (4). The $\alpha$ values range from -0.05 to 0.05 Mm$^{-1}$.

It is important to note that for each case, we have combined IVM data and SOHO/MDI data in order to enlarge the field-of-view and to have a weak magnetic field outside the active region. Those composite images are then compatible with the side boundary conditions described in Section 1.

## 3. Effects of current density on the geometry of field lines

### 3.1. Potential vs nlff magnetic Fields

We proceed to a visual inspection of some particular field lines for both the potential and *nlff* field reconstructions.

For AR 8151, the effects of high current density on the different sets of field lines is strong as seen in Fig. 3: the increase of twist and shear inside the configuration modifies the geometry of the field lines. In particular, the increase of shear modifies the angle between the polarity inversion line and the field line at the apex as seen from the S-shaped field line. We especially note that the connectivity of the field lines (location of the footpoints on the photosphere) is different. From the side view, we notice that in the *nlff* case the field lines are at a lower height in the corona confined by the combination of twist and shear and the surrounding potential field.

For AR 8210, the current density is not strong enough to dramatically modify the magnetic configuration and then the *nlff* field resembles closely the potential field as shown in Fig. 4. From the side view, there is little apparent change in the height of the selected field lines and the different flux bundles have similar footpoints. This active region is characterized by its complex topology evidenced by footpoints close to each other in the negative polarity and connected to different positive polarities. No twisted flux bundles have been reconstructed in this magnetic configuration confirming that the observed H$\alpha$ filaments are low-lying in the chromosphere and the corona and the spatial scale of their associated distribution of current density is smaller than the spatial resolution used to compute the 3D field.

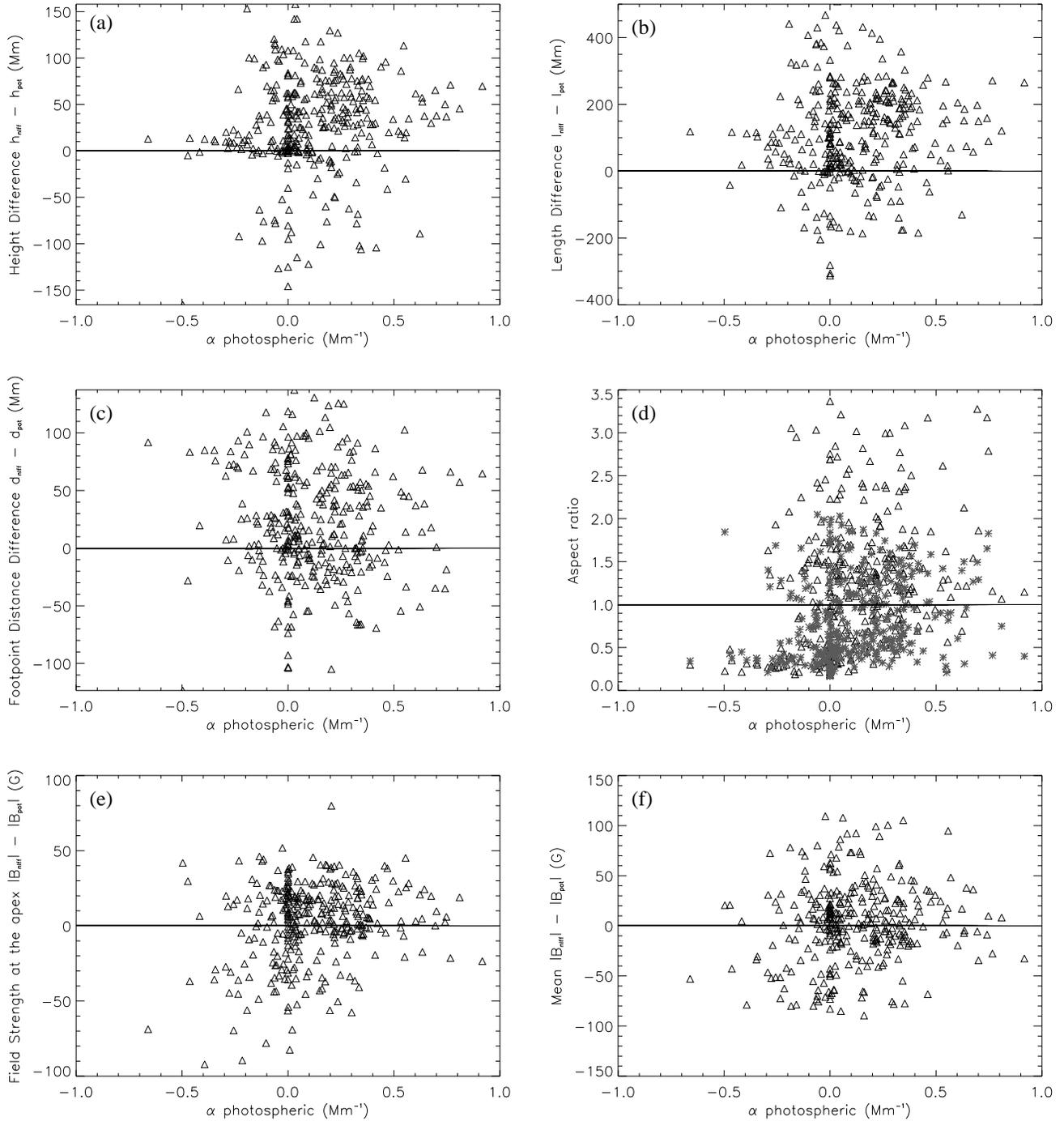

**Fig. 5.** Differences between the potential and *nlff* fields for AR 8151. Geometrical parameters: (a) the height, (b) the length and (c) the distance between footpoints as a function of the $\alpha$ value at the footpoint (in units of Mm$^{-1}$). (d) The aspect ratio of each loop is plotted for the *nlff* (triangles) and for the potential (asterisks). The magnetic field of each loop is characterised by (f) the field strength at the apex and (e) the mean field strength along the loop.

### 3.2. Quantitative Comparison

In order to give a quantitative description of the differences between the potential and *nlff* configurations, we derive geometrical and magnetic parameters for field lines having a field strength at the footpoint above a given threshold ($B_{z,min}$): $h$ the height of the loop (orthogonal projection onto the photo-sphere), $d$ the distance between the two photospheric footpoints of the loop, $l$ the length of the loop, $B_h$ the field strength at the apex of the loop, and $B_{mean}$ the average field strength along the loop. In Figs. 5 and 6, we plot as a function of the photo-spheric values of $\alpha$: (a) the difference in the heights of loops for the potential and *nlff* fields ($h_{nlff} - h_{pot}$), (b) the difference in loop length ($l_{nlff} - l_{pot}$), (c) the difference in the footpoint distance for each loop ($d_{nlff} - d_{pot}$), (d) the aspect ratio, $2h/d$, for both potential (asterisks) and *nlff* (triangles) fields, (e) the difference in magnetic field strength at the apex of each loop ($B_h^{nlff} - B_h^{pot}$), and (f) the difference in the mean magnetic field strength (in absolute value) along each loop ($|B_{mean}^{nlff} - B_{mean}^{pot}|$).

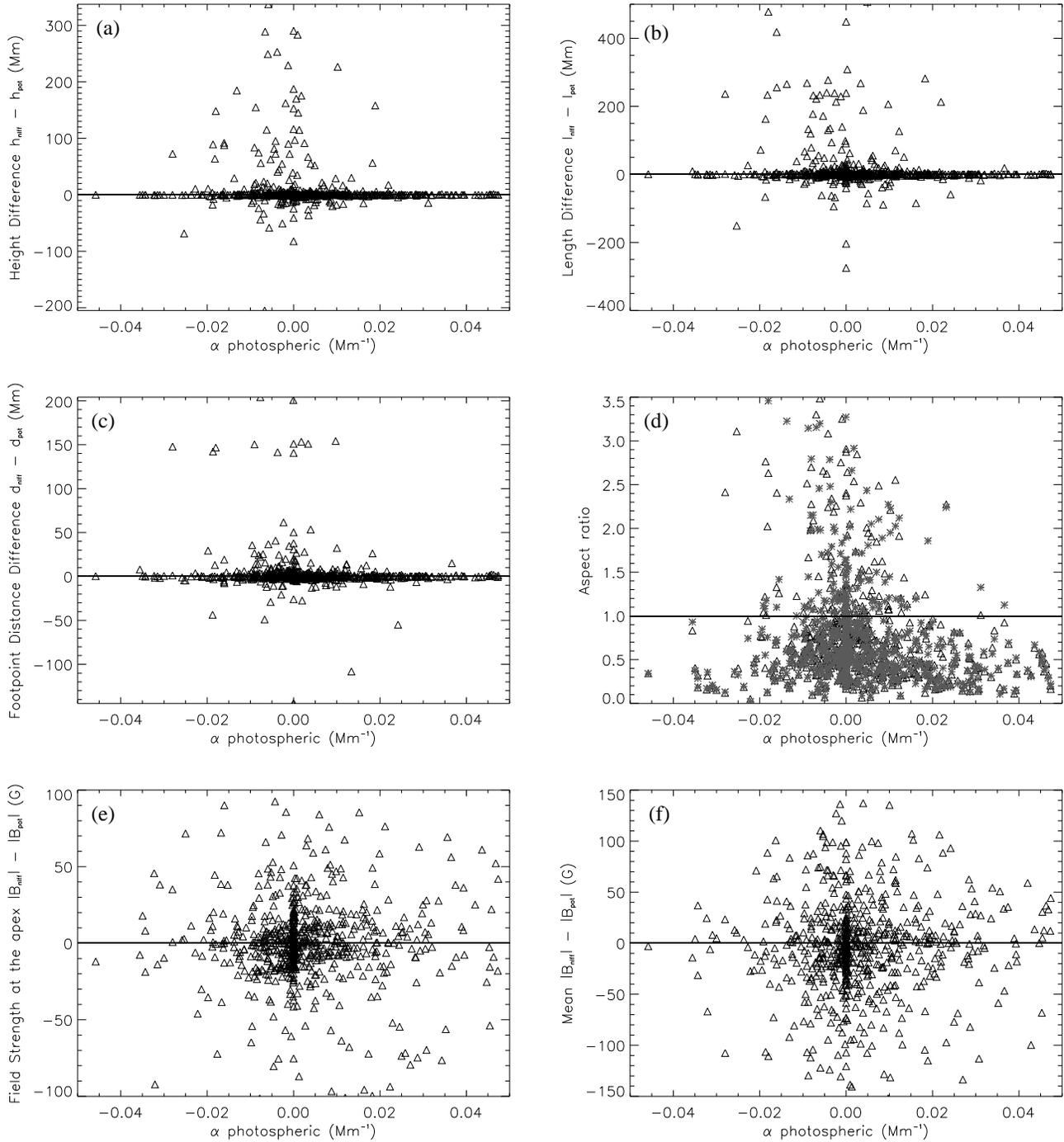

**Fig. 6.** Differences between the potential and *nlff* fields for AR 8210. Geometrical parameters: (a) the height, (b) the length and (c) the distance between footpoints as a function of the $\alpha$ value at the footpoint (in units of Mm$^{-1}$). (d) The aspect ratio of each loop is plotted for the *nlff* (triangles) and for the potential (asterisks). The magnetic field of each loop is characterised by (f) the field strength at the apex and (e) the mean field strength along the loop.

For AR 8151, we have selected $B_{z,min} = 100$ G which limits the number of studied field lines to 552. Note that for this chosen threshold, 38% of the field lines are locally potential ($\alpha = 0$) and so the different parameters are identical for those field lines. In this case with high values of current densities, we calculate the difference between the potential and the *nlff* fields in statistical terms as extracted from Fig. 5. For the geometrical parameters, between 40% and 50% of the *nlff* field lines are higher and longer than potential ones which contradicts our visual inspection described in Sect. 3.1. From Fig. 5a, the mean of the distribution in height is 17.2 Mm with a standard deviation of 43 Mm. From Fig. 5b, the mean of the distribution in length is 68 Mm with a standard deviation of 123.4 Mm. From Fig. 5c, the mean of the distribution in footpoint separation is 13.3 Mm with a standard deviation of 39 Mm. There is a significant change in aspect ratio values from potential to *nlff*: 65% (resp. 35%) of the *nlff* field lines have an aspect ratio less than 1 (resp. greater than 1) and 79% (resp. 21%) of the potential field lines have an aspect ratio less than 1 (resp. greater than 1). In terms of the magnetic field strength, we notice that the values at the apex are statistically higher for the *nlff* field (38% of positive values and 24% of negative values in Fig. 5e) but not significantly with a mean of 1.5 G and a standard deviation of 20.8 G. A similar comment can be made for the mean magnetic field strength along a particular field line with a mean of 4.3 G and a standard deviation of 40.2 G. We can conclude that for AR 8151, the *nlff* field lines are statistically higher, longer and have a stronger magnetic field strength than the potential field lines. An other important point is that the connectivity (parameter *d*) has been significantly modified from one model to the other.

For AR 8210, we choose $B_{z,min} = 100$ G and we will then consider 919 field lines. In this set of field lines, 18% of them are locally potential (with the same parameters). The characteristic parameters are plotted in Fig. 6. 44% (resp. 38%) of the field lines are higher (resp. lower) in the corona for the *nlff* than the potential field with a distribution characterised by a mean of 5.9 Mm and a standard deviation of 34.3 Mm. The results are similar for the length of the loops (47% longer, 35% shorter) and for the distance between the footpoints (45% with an increasing distance and 37% with a decreasing distance). Statistically, the mean values for plots in Fig. 6(a–c) are close to zero and with a small standard deviation with respect to the maximum value of the distribution. We conclude that the geometry and connectivity of the magnetic field lines are similar to those of the potential field lines even if they are carrying current (spreading of $\alpha$ between -0.05 and 0.05 Mm$^{-1}$). This is confirmed by the measurement of the aspect ratio: 84% (resp. 85%) of the *nlff* (resp. potential) field lines have an aspect ratio less than 1. The injection of current density inside the *nlff* magnetic configuration modifies the magnetic strength: both at the apex and on average along the loops, the magnetic field strength is equally distributed around zero with a standard deviation of 28 G at the apex and 43.7 G along the loop for extrema of -100 G and of 100 G. These current carrying field lines can then store magnetic energy.

By comparing quantitatively and statistically the distributions of different characteristic parameters for both active regions, we can conclude that the effects of current density on a magnetic configuration strongly depend on the nature of the magnetic field. For the case of a decaying active region with strong measured electric current density, the magnetic field geometry and connectivity is dramatically modified from the potential field model to the *nlff* model. For the case of the newly emerged active region, the current density along the field lines does not imply strong changes of the magnetic configuration.

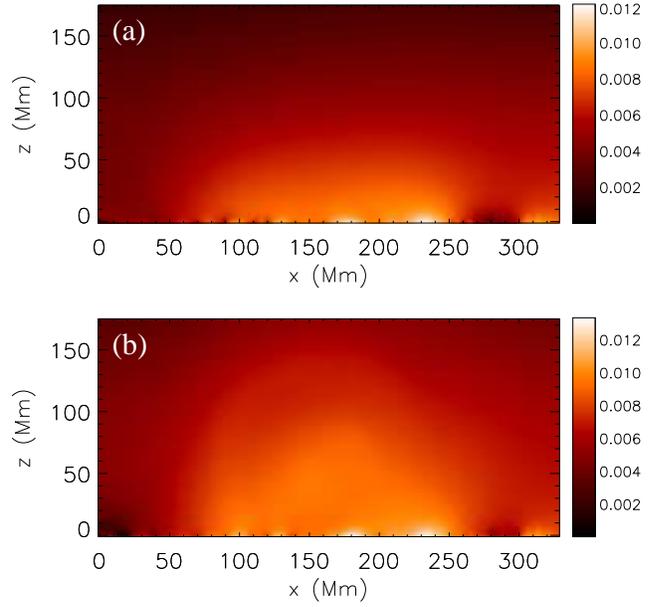

**Fig. 7.** Energy density maps at y = 60 pixel for AR 8151: (a) potential field, (b) *nlff* field (increasing density from black to white). The energy density is in arbitrary units and a log scale is used to take into account the rapid decrease of the magnetic field strength with height.

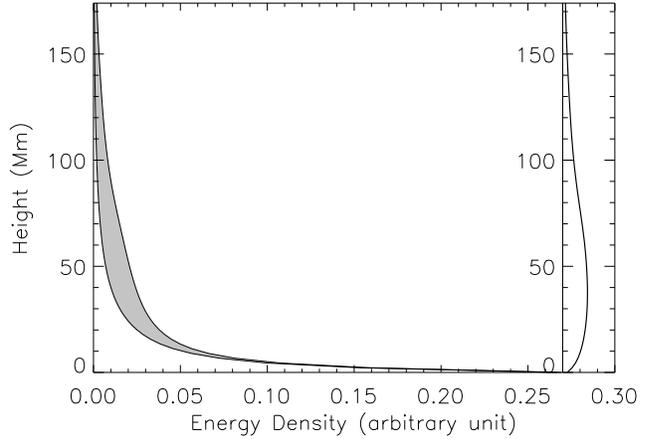

**Fig. 8.** Variation of the energy density for AR 8151 along the *z*-axis for the potential field (lower curve) and the *nlff* field (upper curve) obtained by averaging the magnetic field strength in the corresponding *xy*-plane. The free magnetic energy is contained in the light gray area. The plot on the right shows the percentage of free energy along the *z*-axis (see text for details).

## 4. Comparison of magnetic energy and magnetic helicity

### 4.1. Free Magnetic Energy Budget

One important issue is to know the budget of magnetic energy that can be stored in a magnetic configuration. A part of this energy budget will be released during an eruptive process. The free magnetic energy budget is given by:

$$\Delta E_m = E_m^{nlff} - E_m^{pot} \tag{5}$$

where the magnetic energy $E_m$ is computed in the coronal volume $V$ as follows:

$$E_m = \int_V \frac{B^2}{8\pi}\, dV \qquad (6)$$

**Table 1.** Magnetic energy of the *nlff* field, free magnetic energy budget and relative magnetic helicity for both studied active regions

|  | $E_m^{nlff}$ ($10^{32}$ erg) | $\Delta E_m$ ($10^{32}$ erg) | $\Delta H_m$ ($10^{42}$ Mx$^2$) |
|---|---|---|---|
| AR 8151 | 0.64 | 0.26 | 0.47 |
| AR 8210 | 10.6 | 0.24 | −4.2 |

In addition to the derivation of the above global quantities, we can compute the density of magnetic energy at a given pixel and then determine the location of energy storage in the corona. We study the variations of the energy density with height by averaging the energy density on each *xy*-plane.

For AR 8151, we visualise in Fig. 7 the distribution of magnetic energy density (in arbitrary units) in the plane y = 60 for both the potential and *nlff* fields. We notice that for the *nlff* field magnetic energy is stored in the middle part of the coronal volume. The excess of magnetic energy in the *nlff* configuration is located at the typical heights corresponding to the different twisted flux tubes (Régnier et al. 2002; Régnier & Amari 2004). In Fig. 8, we plot the variation of the magnetic energy density averaged at a given height for both the potential and *nlff* fields. The magnetic energy is mostly located near the photosphere where the magnetic field strength is high but the free magnetic energy (light gray area) is predominantly situated in the middle of the corona in a range of 15 Mm to 70 Mm (as shown by the percentage of free magnetic energy plotted on the right side of Fig. 8). From Table 1, we notice that the free magnetic energy budget is about 40% of the magnetic energy of the whole *nlff* magnetic configuration and is estimated to 2.6 $10^{31}$ erg. This amount of free energy is sufficient to trigger a small energetic flare. Nevertheless the main ingredients which can be responsible for an eruptive event are (i) the existence of highly twisted flux tubes, and (ii) a magnetic energy content close to the Aly-Sturrock limit as discussed in Régnier & Amari (2004).

For AR 8210, the comparison of the deposit of magnetic energy in the corona pictured in Fig. 9 does not show much change from the potential field model to the *nlff* model. The magnetic energy is mainly stored in the low corona close to the photosphere. From Table 1, the free magnetic energy is estimated to 2.4 $10^{31}$ erg corresponding to only 2.5% of the energy of the *nlff* configuration but still enough to trigger small flares. As described in Régnier & Canfield (2006), a series of C-class flares was recorded before and after the particular time studied here. This is also seen in Fig. 10 where we plot the evolution of energy density with height. We have used a logarithmic scale along the energy density axis because both curves are very close from each other. It is noticeable that the percentage of free energy along the *z*-axis shows a concentration of energy in the 50 Mm above the photosphere.

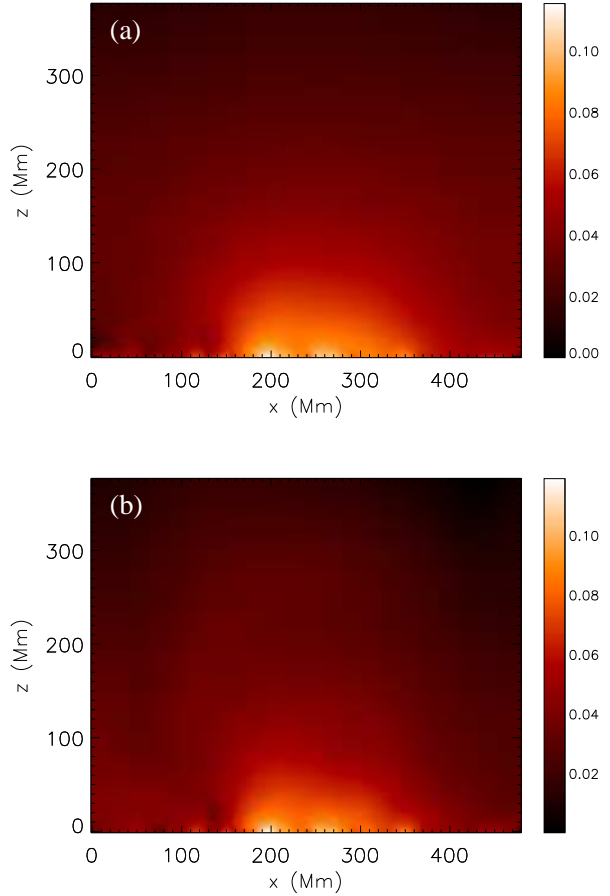

**Fig. 9.** Energy density maps in the plane y = 60 for AR 8210: (a) potential field, (b) *nlff* field (increasing density from black to white). The energy density is in arbitrary units and a log scale is used to take into account the rapid decrease of magnetic field strength with height.

We note that from one active region to the other, the amount of energy in the magnetic configuration can be very different (by a factor of a hundred) depending on the total magnetic flux through the photosphere and on the size of the active region. The percentage of energy stored can also be very different depending on the history or development of the active region prior to the snapshot studied.

### 4.2. Relative Magnetic Helicity

The relative magnetic helicity $\Delta H_m$ is a gauge invariant quantity measuring the twist and shear of a magnetic configuration in a coronal volume $V$ given by (Berger & Field 1984; Finn & Antonsen 1985):

$$\Delta H_m = \int_V (\mathbf{A} - \mathbf{A}_{ref}) \cdot (\mathbf{B} + \mathbf{B}_{ref})\, dV \qquad (7)$$

where $\mathbf{B}$ is the *nlff* field ($\mathbf{B} = \nabla \wedge \mathbf{A}$) and $\mathbf{B}_{ref}$ a reference magnetic field often chosen to be the potential field.

From Table 1, we find that the relative magnetic helicity given by Eqn. (7) is of the order of $10^{42}$ Mx$^2$. The values of the helicity depends on the amount of free energy inside the magnetic configuration as noticed by Kusano et al. (2002). We note that the magnetic helicity for AR 8151 follows the chirality rules defined by Pevtsov et al. (1995) and Longcope et al. (1998): negative helicity sign in the southern hemisphere. This active region follows Joy's law. AR 8210 does not follow either Joy's law or the chirality rules.

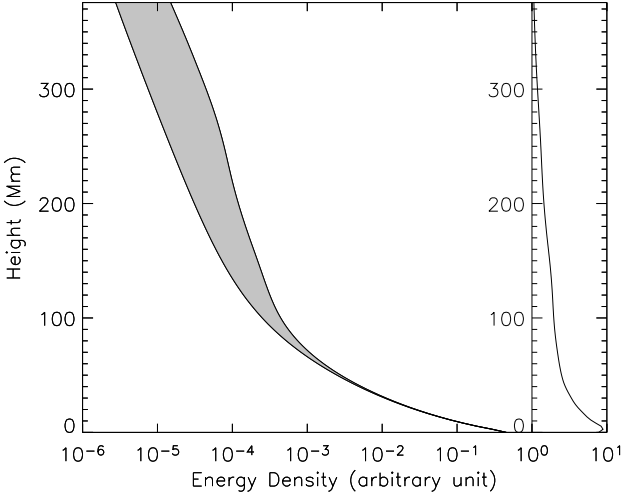

**Fig. 10.** Variation of the energy density for AR 8210 along the *z*-axis for the potential field (lower curve) and the *nlff* field (upper curve) obtained by averaging the magnetic field strength in the corresponding *xy*-plane. The free magnetic energy is contained in the light gray area. The plot on the right shows the percentage of free energy along the *z*-axis (see text for details). Note that the energy density axis is measured on a logarithmic scale.

In a recent review by Démoulin (2006), it is pointed out that for AR 8210 the negative relative magnetic helicity derived from the *nlff* field is not compatible with the sign of helicity derived from the observed H$\alpha$ fibrils in the penumbra of the clockwise rotating sunspot as well as with the positive helicity flux derived by Moon et al. (2002) and Nindos et al. (2003). A positive injection of magnetic helicity can be written as follows:

$$\frac{d \Delta H_m}{dt} > 0 \tag{8}$$

giving for two different times $t_0$ and $t_1 (> t_0)$:

$$\Delta H_m(t_1) - \Delta H_m(t_0) > 0. \tag{9}$$

Taking into account that $\Delta H_m$ can be either positive or negative, we obtain the two following conditions:

$$if \quad \Delta H_m > 0 \quad \Delta H_m(t_1) > \Delta H_m(t_0) \tag{10}$$

$$if \quad \Delta H_m < 0 \quad |\Delta H_m(t_0)| > |\Delta H_m(t_1)|. \tag{11}$$

The condition given by Eqn. (11) agrees with the finding of Régnier et al. (2005) for the relative helicity and of Moon et al. (2002) and Nindos et al. (2003) for the positive injection of helicity flux. Note that Eqn. (11) can also be seen as an annihilation of negative helicity. This injection of flux is mainly dominated by the clockwise rotation of the sunspot. Looking at H$\alpha$ fibrils in the chromosphere is not sufficient to determine the sign of the magnetic helicity of an active region which extends high in the corona and which has a complex distribution of the magnetic field on the photosphere, but it certainly provides a good proxy for the sign of flux injection assuming that sunspot rotation is the main source of injected helicity (see e.g. Démoulin et al. (2002) for a review on the mechanisms of helicity injection due to transverse photospheric motions).

## 5. Conclusions

Our main goal has been to explicitly define the effects of current density on the geometry, connectivity, and energetics of coronal magnetic field configurations. Our first step described in this article is the study of two active regions: a decaying active region with strong current density and a newly emerged active region.

For both examples, the photospheric distributions of current density and of $\alpha$ do not show any particular patterns or any evidence of organised distribution. Nevertheless for the decaying active regions, the $\alpha$ values range from -1 to 1 Mm$^{-1}$ indicating that strong currents are present in the magnetic configuration and are responsible for highly twisted and sheared field lines. While for the newly emerged active region, the $\alpha$ values range between -0.05 and 0.05 Mm$^{-1}$ (4% less than for AR 8151) indicating the existence of weak currents.

From the study of the geometry and the connectivity, we can conclude that the changes due the current density strongly depend on the nature of the active region: the stage of its evolution, the driving velocity field at the bottom boundary responsible for generating currents, the distribution of the sources of magnetic field. For the decaying active region with a simple magnetic distribution, the strong currents generate a twisted flux tube and therefore the departure from the potential field configuration is important. While the weak currents in the newly emerged active region do not dramatically modify the connectivity of the magnetic field lines and the magnetic topology of the configuration.

The strong currents are also responsible for the storage of magnetic energy in the corona ($\sim$ 50 Mm) associated with twisted and sheared flux bundles. The stored magnetic energy is about 40% of the total energy. And for the newly emerged active region, the energy storage is localized close to the photosphere where the magnetic field is stronger and the magnetic field decays with height as fast as a potential field. The stored magnetic energy represents only 2.5% of the total energy but such a value is comparable to the magnetic energy of AR8151, and is sufficient to trigger C- or X-class flares.

*Acknowledgements.* We thank the UK Particle Physics and Astronomy Research Council for financial support (PPARC RG). The computations were done with the XTRAPOL code (supported by the Ecole Polytechnique and the CNES). We also acknowledge the financial support by the European Commission through the SOLAIRE network (MTRN-CT-2006-035484).

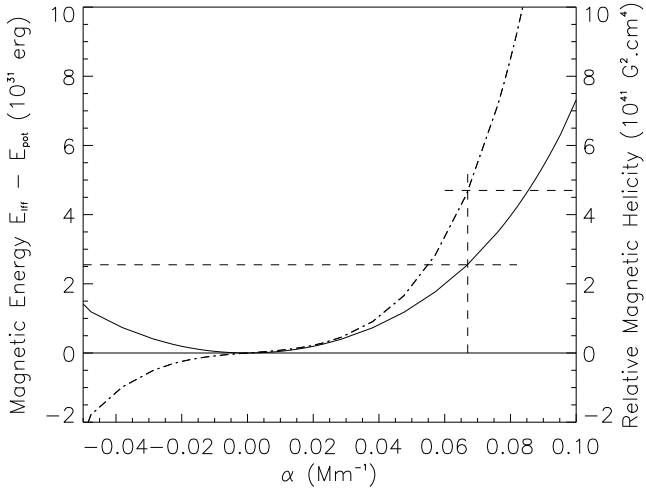

**Fig. 11.** Evolution of the magnetic energy (solid line, unit of $10^{31}$ erg) and magnetic helicity (dot-dashed line, unit of $10^{41}$ G$^2 \cdot$ cm$^4$) as a function of the linear force-free parameter $\alpha$ (Mm$^{-1}$). The dashed lines indicate the $\alpha$ value and the magnetic energy value for the *lff* having the same magnetic helicity as the *nlff* field.

## Appendix A: Comparison of magnetic helicity values

In this article, we compared the *nlff* fields with the potential fields but deliberately side-stepped a comparison with a linear force-free field. Even if the computation of a linear force-free field is easier and faster, we justify this decision as follows:

(i) observations of the transverse field on the photosphere reveal a highly non-uniform distribution of $\alpha$;
(ii) the best choice of the $\alpha$ value is somehow arbitrary and is still debated (see e.g., Leka & Skumanich 1999; Leka et al. 2005);
(iii) the linear force-free corresponds to a mathematically ill-posed problem. For instance, Seehafer (1978) has demonstrated that the magnetic energy in an infinite volume cannot be bounded;
(iv) the magnetic helicity content derived from the *lff* field does not measure the complexity of the field as observed in the corona. In particular in a single active region both signs of helicity (e.g., right and left handed flux tubes) can be found.

Therefore the *lff* approximation does not contain important physical ingredients that are possible for a *nlff* field, such as twisted bundles or highly sheared arcades. Direct comparisons between measurements in the low corona and different models have highlighted the additional physics contained in the *nlff* approximation (e.g. Wiegelmann et al. 2005). Nevertheless, the *lff* field having a magnetic helicity $\Delta H_m^{lff}$ does represent a minimum energy state of the *nlff* field having the same magnetic helicity $\Delta H_m^{nlff} = \Delta H_m^{lff}$ (Molodensky 1969; Aly 1989). Therefore, it is interesting to compare the magnetic energy and magnetic helicity contents for *lff* field with various $\alpha$ values and the *nlff* field. We perform this experiment for AR 8151 because of the existence of flux tubes with opposite handedness.

We compute the *lff* field using the Grad-Rubin numerical scheme with the same boundary conditions as for the *nlff* field and for a range of $\alpha$ values from −0.05 to 0.1. We then derive the magnetic energy given by Eqn. (6) and the relative magnetic helicity given by Eqn. (7). The results are presented in Fig. 11. We notice that the free magnetic energy in a *lff* configuration (solid line) evolves as $\alpha^2$ and that the magnetic helicity (dot–dashed line) evolves as $\alpha^3$. From Fig. 11, we take the value of the *nlff* magnetic helicity (dashed line at $\Delta H_m = 4.7\ 10^{41}$ G$^2 \cdot$ cm$^4$), we deduce the associated value of $\alpha$ for the *lff* field (dashed line at $\alpha = 6.7\ 10^{-2}$ Mm$^{-1}$) and then we find that the corresponding magnetic energy budget is of $\Delta E_m^{lff} = 2.55\ 10^{31}$ erg. The latter value gives a magnetic energy for this particular *lff* field of $6.55\ 10^{31}$ erg comparable to the magnetic energy of the *nlff* field (see Table 1). The *lff* magnetic configuration with this magnetic energy corresponds to a minimum energy state of the *nlff* field. In terms of geometry, the *lff* magnetic field configuration corresponding to this minimum energy state has a different connectivity but is closer to the *nlff* field than the potential field configuration, and we were not able to recover any twisted flux tubes in the 3D configuration.

It is important to note that the mean value of $\alpha$ on the photosphere is negative giving a negative magnetic helicity opposite to our results derived from the *nlff*. Therefore when performing a *lff* field reconstruction, the choice of $\alpha$ is crucial and can lead to wrong conclusions even in the sign of the magnetic helicity.

## References

Alissandrakis, C. E. 1981, A&A, 100, 197
Altschuler, M. D. & Newkirk, Jr., G. 1969, Sol. Phys., 9, 131
Aly, J. J. 1989, Solar Phys., 120, 19
Amari, T., Aly, J. J., Luciani, J. F., Boulmezaoud, T. Z., & Mikic, Z. 1997, Solar Phys., 174, 129
Amari, T., Boulmezaoud, T. Z., & Aly, J. J. 2006, A&A, 446, 691
Amari, T., Boulmezaoud, T. Z., & Mikic, Z. 1999, A&A, 350, 1051
Berger, M. A. & Field, G. B. 1984, Journal of Fluid Mechanics, 147, 133
Bleybel, A., Amari, T., van Driel-Gesztelyi, L., & Leka, K. D. 2002, A&A, 395, 685
Canfield, R. C., Hudson, H. S., & McKenzie, D. E. 1999, Geophys. Res. Lett., 26, 627
Chiu, Y. T. & Hilton, H. H. 1977, ApJ, 212, 873
Close, R. M., Heyvaerts, J. F., & Priest, E. R. 2004, Sol. Phys., 225, 267
Démoulin, P., Mandrini, C. H., Van Driel-Gesztelyi, L., Lopez Fuentes, M. C., & Aulanier, G. 2002, Solar Phys., 207, 87
Démoulin, P. 2006, Advances in Space Research, submitted
Finn, J. M. & Antonsen, T. M. 1985, Comments Plasma Phys. Controlled Fusion, 9, 111
Gary, G. A. 1989, ApJ, 69, 323
Grad, H. & Rubin, H. 1958, in Proc. 2nd Int. Conf. on Peaceful Uses of Atomic Energy, Geneva, United Nations, Vol. 31, 190
Jiao, L., McClymont, A. N., & Mikic, Z. 1997, Sol. Phys., 174, 311

Krall, K. R., Smith, Jr., J. B., Hagyard, M. J., West, E. A., & Cummings, N. P. 1982, Sol. Phys., 79, 59
Kusano, K., Maeshiro, T., Yokoyama, T., & Sakurai, T. 2002, ApJ, 577, 501
Leka, K. D., Fan, Y., & Barnes, G. 2005, ApJ, 626, 1091
Leka, K. D. & Skumanich, A. 1999, Solar Phys., 188, 3
Longcope, D. W., Fisher, G. H., & Pevtsov, A. A. 1998, ApJ, 507, 417
Mackay, D. H. & van Ballegooijen, A. A. 2006, ApJ, 641, 577
Maclean, R. C., Beveridge, C., Hornig, G., & Priest, E. R. 2006a, Sol. Phys., 237, 227
Maclean, R. C., Beveridge, C., & Priest, E. R. 2006b, Sol. Phys., 238, 13
Martin, S. F. 1998, Solar Phys., 182, 107
Masuda, S., Kosugi, T., Hara, H., et al. 1995, PASJ, 47, 677
Metcalf, T. R., Jiao, L., McClymont, A. N., Canfield, R. C., & Uitenbroek, H. 1995, ApJ, 439, 474
Mickey, D. L., Canfield, R. C., Labonte, B. J., et al. 1996, Solar Phys., 168, 229
Molodensky, M. M. 1969, Soviet Astron. - AJ,, 12, 585
Moon, Y.-J., Choe, G. S., Yun, H. S., Park, Y. D., & Mickey, D. L. 2002, ApJ, 568, 422
Nakagawa, Y. & Raadu, M. A. 1972, Solar Phys., 25, 127
Nindos, A., Zhang, J., & Zhang, H. 2003, ApJ, 594, 1033
Pevtsov, A. A., Canfield, R. C., & Metcalf, T. R. 1995, ApJ, 440, L109
Priest, E. R. & Forbes, T. G. 2002, A&A Rev., 10, 313
Priest, E. R., Hood, A. W., & Anzer, U. 1989, ApJ, 344, 1010
Rayrole, J. & Semel, M. 1970, A&A, 6, 288
Régnier, S. 2007, in Mem. S.A.It.
Régnier, S. & Amari, T. 2004, A&A, 425, 345
Régnier, S., Amari, T., & Canfield, R. C. 2005, A&A, 442, 345
Régnier, S., Amari, T., & Kersalé, E. 2002, A&A, 392, 1119
Régnier, S. & Canfield, R. C. 2006, A&A, 451, 319
Riley, P., Linker, J. A., Mikić, Z., et al. 2006, ApJ, 653, 1510
Rust, D. M. & Kumar, A. 1996, ApJ, 464, L199
Sakurai, T. 1981, Sol. Phys., 69, 343
Schmidt, H. U. 1964, in The Physics of Solar Flares, ed. W. N. Hess, p.107
Schrijver, C. J., Derosa, M. L., Metcalf, T. R., et al. 2006, Solar Phys., 235, 161
Seehafer, N. 1978, Sol. Phys., 58, 215
Semel, M. 1967, Annales d'Astrophysique, 30, 513
Wiegelmann, T. 2004, Sol. Phys., 219, 87
Wiegelmann, T., Inhester, B., & Sakurai, T. 2006, Sol. Phys., 233, 215
Wiegelmann, T., Lagg, A., Solanki, S. K., Inhester, B., & Woch, J. 2005, A&A, 433, 701
Wiegelmann, T. & Neukirch, T. 2006, A&A, 457, 1053